\documentclass[journal]{IEEEtran}
\usepackage{times,amsmath,color,amssymb,graphicx,epsfig,cite,psfrag,subfigure,algorithm,balance}
\usepackage{amsfonts,pifont,enumerate,cases}
\usepackage{mathrsfs} 
\usepackage[table]{xcolor} 
\usepackage{verbatim} 
\usepackage{bm}

\newtheorem{remark}{\underline{Remark}}

\newcounter{MYtempeqncnt}

\begin{document}

\title{ Beam Tracking for UAV Mounted SatCom on-the-Move with  Massive  Antenna  Array}
\author{Jianwei Zhao, 
Feifei Gao, Qihui Wu, Shi Jin, Yi Wu, 
Weimin Jia
\thanks{J. Zhao and F. Gao are with  Tsinghua National Laboratory for Information Science and Technology (TNList) Beijing 100084, P. R. China (e-mail: zhaojw15@mails.tsinghua.edu.cn, feifeigao@ieee.org). J. Zhao is also with
High-Tech Institute of Xi’an, Xi’an, Shaanxi 710025, China. }
\thanks{Q. Wu is with the Department of Electronics and Information Engineering,
Nanjing University of Aeronautics and Astronautics, Nanjing 210007, China
(email: wuqihui2014@sina.com).}
\thanks{S. Jin is with the National Communications Research Laboratory, Southeast University, Nanjing 210096, P. R. China (email: jinshi@seu.edu.cn). }
\thanks{Yi Wu is with College of Photonic and Electronic Engineering, Fujian Normal University, Fuzhou, 350117, China
(e-mail: wuyi@fjnu.edu.cn).}
\thanks{W. Jia is with High-Tech Institute of Xi’an, Xi’an, Shaanxi 710025, China
(e-mail: jwm602@163.com).}
}
\maketitle
\thispagestyle{empty}

\begin{abstract}
Unmanned aerial vehicle (UAV)-satellite communication has drawn dramatic attention for its potential to build the integrated space-air-ground network and the  seamless wide-area coverage. The key challenge to UAV-satellite communication is its unstable beam pointing  due to the UAV navigation, which is a typical \emph{SatCom on-the-move} scenario. In this paper, we propose a blind beam tracking approach for Ka-band UAV-satellite communication system, where UAV  is equipped with a large-scale antenna array. The effects of UAV navigation are firstly released through the mechanical adjustment, which could approximately point the beam towards the target satellite through  \emph{beam stabilization} and  \emph{dynamic isolation}. Specially, the attitude information can be realtimely derived from data fusion of low-cost sensors. Then, the precision of the beam pointing is blindly refined through  electrically adjusting the weight of the massive antennas, where an array structure based \emph{simultaneous perturbation} algorithm is designed. Simulation results are provided to demonstrate the superiority of
the proposed method over the existing ones.
\end{abstract}

\begin{IEEEkeywords}
UAV, SatCom on-the-move, beam tracking, mechanical adjustment, electrical adjustment, massive antenna array.
\end{IEEEkeywords}

\section{Introduction}
\begin{figure}[t]
\centering
\includegraphics[width=90mm]{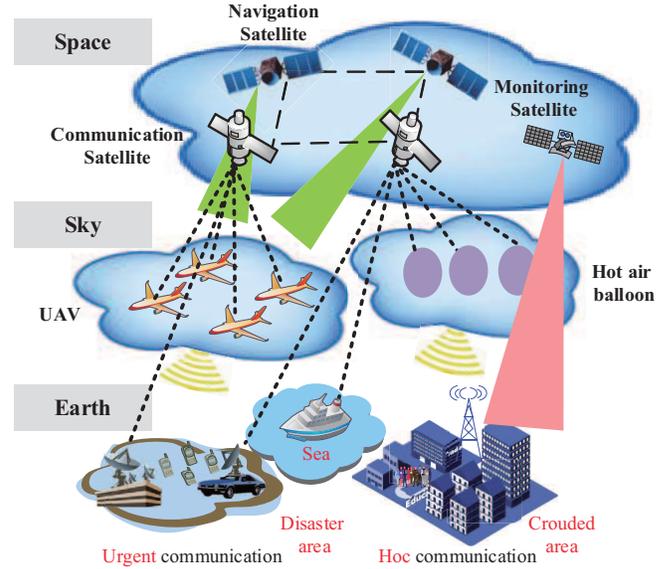}
\caption{The integrated space-air-ground network.}
\label{fig:systemfig}
\end{figure}

Satellite communication has become a promising solution to realize high data rates anywhere, anytime, and towards the seamless wide-area coverage , in which unmanned aerial vehicle (UAV)-satellite communication\footnote{Geosynchronous satellite is usually used for UAV-satellite communication since  its location relative to earth keeps invariant.} is a key part for building the integrated space-air-ground network \cite{uav1,uav2}. As is shown in Fig. \ref{fig:systemfig}, UAV communication could be adopted to serve users without infrastructure coverage such as disaster areas after earthquake, where the ground communication infrastructure is destroyed, or could assist the existing communications, e.g., the rapid service recovery and the base station (BS) offloading of the extremely crowded areas. In these circumstances, UAV acts as a relay,  connecting to terrestrial networks via a satellite link, and serves user terminals via a ground link \cite{trajectory}.

On the other side, spectrum crowding and increased data rates encourage the migration to  millimeter wave (mmWave) band, typically Ka-band \cite{mmwave1,5db,mmwave2,mmwave3,mmwave4}.  Though the heavy path loss of Ka-band seemingly blocks it from practical usage, the micro wavelength of Ka-band permits the employment of large amount of antennas onto a small area  to achieve adequate spatial gains and combat such propagation losses \cite{overview,CCS,precoding,summary6,massivemimo,five}. Meanwhile, multiple-antennas with small size would be specifically suitable for UAV mounted transmission, which  also removes the requirement of the expensive high-gain directional antenna in satellite communication and thus greatly reduces the overall system cost.

In general, the performances of the communication systems critically rely on the channel state information (CSI), which brings a large number of CSI estimation works \cite{xie,xieCR,gao1,gao2,gao3,xies,dai,channelestimation}. Different from the conventional wireless communications systems, UAV based one would encounter new challenges. Due to the continuous navigation of UAV, the channel would vary constantly for both the UAV-satellite link and the UAV-ground link. Furthermore, because of the constraints of the overall system cost and the power consumption, the available radio frequency (RF) chains are limited for UAV-satellite communication systems. A beam sweeping method was adopted in~\cite{channelestimation} for beam tracking, which spends significant resources to train the beam in all directions. A priori knowledge aided channel tracking method was proposed in \cite{dai} for Teraherz massive MIMO system, where a linear motion is adopted to derive angles of the incident signals and then the complex gains are estimated by pilots. However, the linear movement model restricts its application for   practical nonlinear movement scenes. An angle division multiple access (ADMA) based channel tracking scheme was proposed in \cite{xies} for massive MIMO systems, where channel tracking is transmitted to estimating the direction of arrival (DOA) information of the incoming signals and the time-varying complex gain information, which greatly reduces the complexity  of channel tracking. 

Albeit  most  methods   \cite{xie,xieCR,gao1,gao2,gao3,xies,dai,channelestimation} can be extended to the UAV-ground link, the channel tracking for UAV-satellite link still needs to be  investigated due to its distinct properties. For UAV-satellite link, the premise of establishing a successful link is the alignment of the spatial beam from UAV to satellite, i.e., the spatial beam should keep ``locking"   the direction of the target satellite during UAV navigation \cite{range,channelmode2,beamalignment,wu}, which is a typical SatCom on-the-move. However, the attitude variation of UAV (including yaw angle, pitch angle and roll angle) would constantly affect the beam pointing. Therefore, UAV needs to overcome the effects of its navigation, and persistently adjusts the spatial beam towards the target satellite. 
 An electrical tracking method for phased-array antenna system was developed in \cite{beamalignment}, which adopts the electrical feedback to track the beam pointing variation. However, for UAV-satellite link, it is always preferred that the steered spatial beam should be pointed from the normal of the array antenna plane since the array antenna gain is proportional to the projected area of the antenna aperture in the direction of the target satellite~\cite{range}, which demands for the mechanical adjustment as did in conventional SatCom on-the-move~\cite{wu}.  

 In this paper, we propose a blind hybrid beam tracking method for the UAV-satellite communication with a massive uniform plane array (UPA). 
The mechanical adjustment is firstly adopted to alleviate the effects of UAV navigations, where the spatial beam can be pointed towards the target satellite through the \emph{beam stabilization} algorithm and \emph{dynamic isolation} algorithm. However, due to the limited precision of low cost attitude sensors, the system error, and measurement error, the mechanical adjustment could only roughly  points the spatial beam of UAV  to the target satellite. We then design an electrical adjustment to further calibrate beam pointing, where the spatial beam is steered to DOA of the incident signals in a more accurate manner with an array structure based \emph{simultaneous perturbation}  algorithm.  Finally, simulation results are provided to verify the effectiveness of the proposed method and demonstrate its superiority over the existing candidates.

The rest of the paper is organized as follows. The system model is described in section \ref{sec:model}. Beam tracking with mechanical adjustment is presented in section \ref{sec:attitude}, while beam tracking  with electrical adjustment is presented in section \ref{sec:simultaneous}. The simulations are then displayed in section~\ref{sec:simulation}. Finally, conclusions are drawn in section \ref{sec:conclusions}.

\textbf{Notations:} Vectors and matrices are denoted by boldface small and
capital letters;  the transpose, complex conjugate, Hermitian,
inverse, and pseudo-inverse of the matrix ${\mathbf A}$ are denoted
by ${\mathbf A}^T$, ${\mathbf A}^*$, ${\mathbf A}^H$, ${\mathbf
A}^{-1}$ and ${\mathbf A}^\dag$, respectively.

\section{Problem Formulation}\label{sec:model}
\subsection{Channel Model}
Let us consider a UAV-satellite communication system, where UAV  is equipped with UPA of $M\times N$ antennas with $M\gg1, N\gg1$, while a single 
antenna is installed on the geosynchronous satellite \cite{satellite1,satellite2}. Due to the constraints of system cost and power consumption, the available radio frequency (RF) chains are limited for UAV-satellite communication system. In this paper, it is assumed that UAV has only one dedicated RF chain, i.e., single channel receiver mode~\cite{RF1,RF2}. 

Define $\mathbf{A}(\alpha_{l},\beta_{l})\in \mathbb{C}^{M\times N}$ as the array response matrix, whose $(m,n)$-th element can be expressed as
\begin{align}\label{a2}
\mathbf{A}(\alpha_{l},\beta_{l})=e^{j\frac{2\pi d}{\lambda_c}\left[(m-1)\sin \alpha_{l} \cos \beta_{l}+(n-1)\sin \alpha_{l} \sin \beta_{l}\right]},
\end{align}
where $\alpha_{l}$ and $\beta_{l}$ are the azimuth angle and elevation angle of the $l$-th path from the target satellite.
Then, the $M\times N$ channel matrix $\mathbf{H}$ between the target satellite and UAV can be expressed as~\cite{xie,channelmode2}
\begin{align}\label{equ:channelmodel}
\mathbf{H}=\sum_{l=1}^{L}\frac{a_{l}e^{-j2\pi d_{l}/\lambda_c}}{\sqrt{MN}} \mathbf{A}(\alpha_{l},\beta_{l}),
\end{align}
where $L$ represents the number of the incident pathes, $d_{l}$ is the distance between antenna 1 of UAV and the satellite along path $l$, and $\lambda_c$ is the carrier wavelength. Moreover, $a_{l}$ is the complex channel attenuation of the $l$-th path due to gaseous, cloud, rain, and scintillation attenuation. A combination method considering the above effects can be seen in \cite{channelmode2}.

\begin{figure}[t]
\centering
\includegraphics[width=90mm]{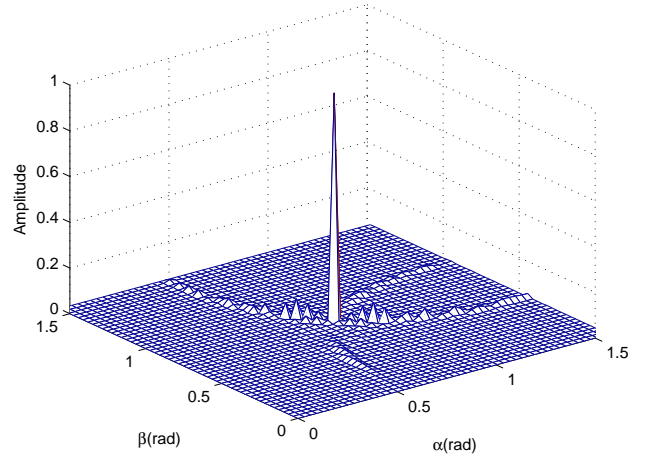}
\caption{The characteristic of Ka-band massive array antenna.}
\label{fig:channelcha}
\end{figure}

For UAV-satellite link, there are limited scatters in the propagation surroundings and thus the number of the incident paths meets $L\ll N$ and $L\ll M$. Propagation measurement says that the overwhelming majority of the total energy arrives at the receiver from the  direct path in satellite communication~\cite{range}. Meanwhile,  fixed satellite communication systems above 10 GHz have been verified to operate under loss of sight (LOS) only \cite{channelmode2}. Therefore, LOS ray is the dominant one while the other Non-LOS (NLOS) components can be ignored for the Ka-band UAV-satellite communication link.

Define the spatial domain channel matrix as the 2-dimension discrete Fourier transform (2D-DFT) of $\mathbf H$, i.e.,
\begin{align}\label{equ:channelmodelDFT}
\tilde {\mathbf H}=\mathbf{F}_M \mathbf H\mathbf{F}_N,
\end{align}
where  $\mathbf{F}_M $ and $
\mathbf{F}_N $ are   the normalized  DFT matrices with dimension $M\times M$ and $N\times N$. According to the analysis in \cite{xie}, the significant values of the spatial channel $\tilde {\mathbf H}$ are centered around DOA $\alpha$ and $\beta$ of the impinging signal, as shown in Fig.~\ref{fig:channelcha}. Besides, the massive number of antennas could form narrow beam with high spatial gain, and thus is very suitable for UAV-satellite communication.

\subsection{Problem Formulation}
\begin{figure}[t]
\centering
\includegraphics[width=85mm]{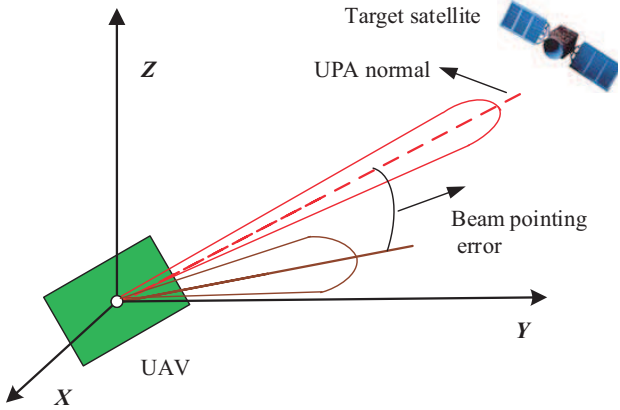}
\caption{The error analysis of UAV beam pointing, where the beam direction should be stabilized at the vicinity of the plane normal for high quality of communication.}
\label{fig:communication_strategy}
\end{figure}

For UAV-satellite link, the beam pointing error would severely degrade the communication quality, which should be kept as small as possible. In addition, it has been verified that the optimal working range of UPA is in the vicinity of its normal, and the array  gain is proportional to the projected area of the antenna aperture in the direction of the target satellite, which declines by sine of the elevation angle \cite{range}. For example in Fig.~\ref{fig:communication_strategy}, the spatial beam would be distorted if it deviates far from the direction of the plane normal.

In general, there are two main factors  affecting beam pointing: satellite perturbation and UAV navigation.
 In order to overcome the influence of the satellite perturbation, the position-keeping technique is usually adopted so that the latitude and longitude errors of the satellite position are kept within the allowable range.  A good example of  position-keeping technique  is \cite{satellite-keeping}, where satellite perturbation has little effect on beam pointing within one hour and can be neglected. On the other side, the UAV navigation,  including the carrier's heading, pitching and rolling motion, are very detrimental. Unfortunately, there is no corresponding method to overcome the effects of UAV navigation for UAV-satellite communication systems.

\begin{figure}[t]
\centering
\includegraphics[width=70mm]{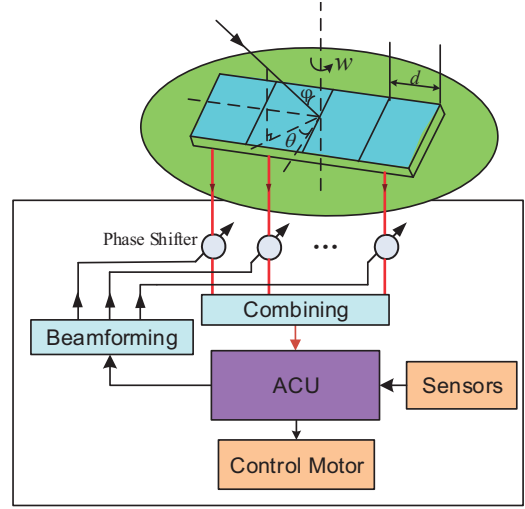}
\caption{The structure of the UAV communication system, where a mechanical adjustment subsystem is installed at BS and each antenna is equipped with a analog shifters instead of a dedicated RF chain.}
\label{fig:structure}
\end{figure}

To remedy the effects of UAV navigation, we propose a new  beam tracking method based on the joint  mechanical adjustment and electrical adjustment. The structure of the designed beam tracking system is depicted in Fig. \ref{fig:structure}, which consists of three segments: (1) the perception part  contains the measurement sensors and is responsible to sense the UAV attitude variations as well as to measure the corresponding angular rate; (2) the decision part, i.e., the arithmetic and control unit (ACU),  provides control commands to stabilize beam pointing according to the outputs of the perception part;~(3) the operative part  contains the control motors (the azimuth angle control monitor, the polarization angle control monitor, and the elevation angle control monitor) and phase shifters, etc., which implements the directives of the decision part.

Traditionally, the high precision inertial navigation system~\cite{wu} is used to obtain the attitude information in UAV military applications. However, the cost is not affordable for the civil applications. Recently, with the development of the micro electro mechanical system (MEMS), there appear many low-cost attitude sensors, such as MEMS gyros, MEMS accelerator, global position system (GPS). Therefore, we propose to utilize these low-cost sensors to derive the attitude information and reduce the total system cost. The attitude determination principles of these low-cost sensors can be found  in   Appendix~A.

\section{Coarse Beam Alignment with Mechanical Adjustment}\label{sec:attitude}
\begin{figure}[t]
\centering
\includegraphics[width=90mm]{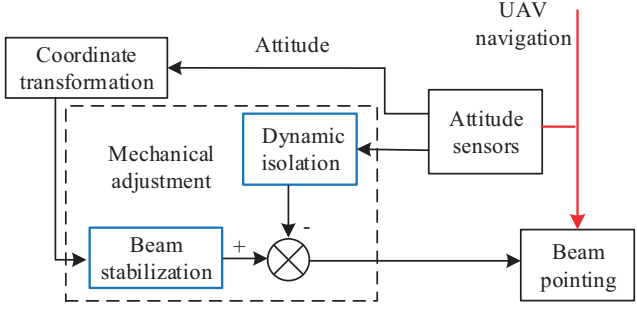}
\caption{Mechanical adjustment with beam stabilization and dynamic isolation.}
\label{fig:tracking_scheme}
\end{figure}

Due to small wavelength of Ka-band, the size of massive array antenna is not large, which makes the mechanical adjustment based beam tracking applicable \cite{wu}. In general, the mechanical adjustment for beam tracking consists of two steps: the \emph{beam stabilization} and the \emph{dynamic isolation}, which  is shown in Fig.~\ref{fig:tracking_scheme}.

\subsection{Beam Stabilization}
The proposed beam stabilization  jointly utilizes the UAV attitude information and satellite location information to track the spatial beam. To represent UAV attitude information, we need to explain  the reference frames first, which describe the relationships between UAV attitude and beam pointing. Different reference frames are  also shown in Fig.~\ref{fig:frame}.
\begin{enumerate}

\item The geodetic coordinate frame ($n$-frame): Its origin is chosen as the UAV center of gravity (UAVCG), and its three axes $x_n,\ y_n,\  z_n$ are separately aligned with the direction of north, east, and geocentric.

\item The UAV-fixed coordinate frame ($b$-frame): Its origin is UAVCG, and its three axes $x_b,\ y_b,\ z_b$, i.e., roll axis, pitch axis, and yaw axis, are aligned with its longitudinal (forward), lateral (right), and vertical (downward) direction.

\item The azimuth-rotary coordinate frame ($a$-frame): Its origin is the array antenna center of gravity (AACG). Its axis $z_a$ is parallel to $z_b$, while axis $y_a$ is identical with the direction of UPA elevation axis. The axis $x_a$ is perpendicular to the plane spanned by the axis $y_a$ and $z_a$. According to the relationship of $b$-frame and $a$-frame, the $b$-frame would be consistent with the $a$-frame when we rotate the azimuth angle $\alpha$ around the axis $z_b$. 

\begin{figure}[t]
\centering
\includegraphics[width=80mm]{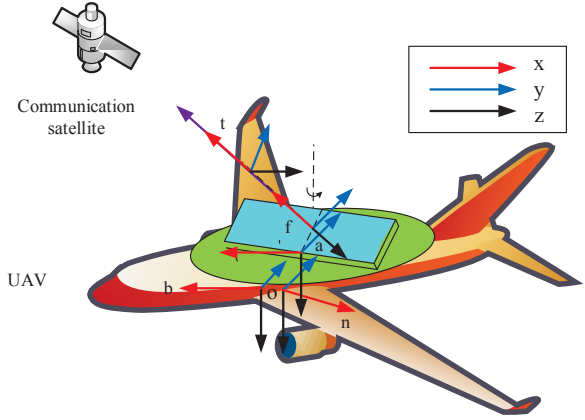}
\caption{The definition of different reference frames, including $n$-frame, $b$-frame, $a$-frame, $f$-frame and $t$-frame.}
\label{fig:frame}
\end{figure}

\item The UPA coordinate frame ($f$-frame): Its origin is AACG. Besides, its axis $y_f$ is parallel to $y_a$, while its  $x_f$ axis points to the satellite.  The axis $z_f$ is perpendicular to the plane spanned by the axis $x_f$ and $y_f$. On the basis of the relationship of $f$-frame and $a$-frame, the $a$-frame would be consistent with the $f$-frame when we rotate the elevation angle $\beta$ around the axis $y_a$. 

\item The spatial beam coordinate frame ($t$-frame): Its origin is also AACG, and axis $y_t$ is identical with the direction of electric field, while $x_t$  is parallel to $x_f$.  The axis $z_t$ is perpendicular to the plane spanned by the axis $x_t$ and $y_t$. Based on the relationship between the $f$-frame and $t$-frame, the $f$-frame would be consistent with the $t$-frame when we rotate the polarization angle $\gamma$ around the axis $x_t$. 
\end{enumerate}

\begin{figure}[t]
\centering
\includegraphics[width=80mm]{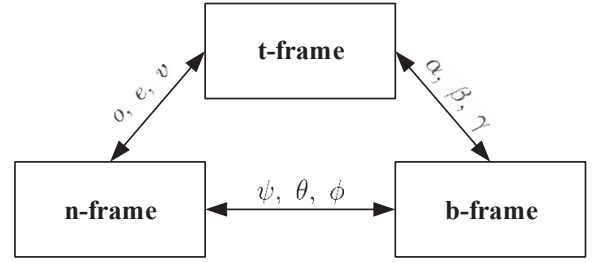}
\caption{The relationships between different frames.}
\label{fig:frame_relaltionship}
\end{figure}

The UAV attitude relative to the $n$-frame is determined by the yaw angle $\Psi$, the pitch angle $\theta$, and the roll angle $\phi$, while the beam attitude of UAV array antenna relative to the $b$-frame is dependent on the azimuth angle $\alpha$, the elevation angle $\beta$, and the polarization angle $\gamma$. Moreover, the beam attitude of UAV array antenna relative to the $n$-frame can be expressed by the Euler angles $o$, $e$, and $v$. The relationships between different frames are shown in Fig. \ref{fig:frame_relaltionship}, and the transformation between different frames can be realized by the coordinate transformation matrix~\cite{wu}.

 Denote $\mathbf C_b^t$ as the coordinates transformation matrix between the $b$-frame and the $t$-frame, which can be realized by three continuous spatial rotation: rotating $\alpha$ around the axis $z_b$ that is represented by ${\mathbf T_1}(\alpha) \!\!= \!\!\!\left[ \!\!{\begin{array}{*{20}{c}}
{\cos \alpha }&{\sin \alpha }&0\\
{ - \sin \alpha }&{\cos \alpha }&0\\
0&0&1
\end{array}} \!\!\right]$, rotating $\beta$ around the axis $y_a$ that is represented by ${\mathbf T_2}(\beta ) \!\!= \!\!\!\left[\!\! \!{\begin{array}{*{20}{c}}
{\cos \beta}&0&{ - \sin \beta }\\
0&1&0\\
{\sin \beta }&0&{\cos \beta }
\end{array}}\! \!\!\right]$, and rotating $\gamma$ around the axis $x_t$ that is represented by ${\mathbf T_3}(\gamma ) \!\!= \!\!\!\left[\! \!{\begin{array}{*{20}{c}}
1&0&0\\
0&{\cos \gamma }&{\sin \gamma }\\
0&{ - \sin \gamma }&{\cos \gamma }
\end{array}} \!\right]$.

\begin{figure*}[t]
\centering
\includegraphics[width=140mm]{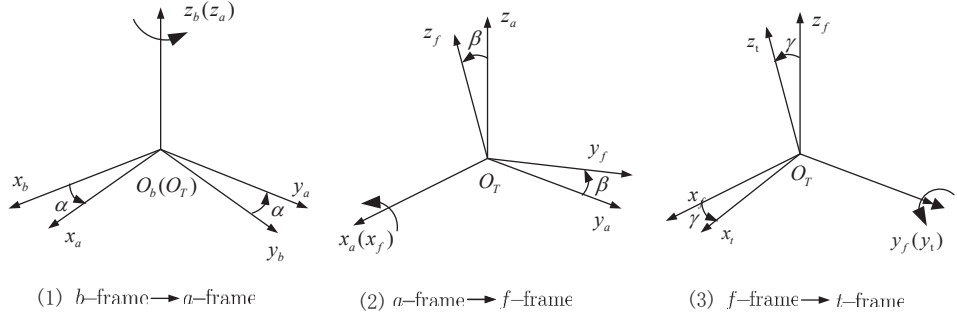}
\caption{The transformation from the $b$-frame to $t$-frame.}
\label{fig:transformation}
\end{figure*}

 Each of the three rotations are illustrated in Fig. \ref {fig:transformation}. Thus, $\mathbf C_b^t$ can be mathematically  expressed as \eqref{equ:transform2}, shown on the top of the next page.
 \begin{figure*}[!t]
\normalsize
\setcounter{MYtempeqncnt}{\value{equation}}
\setcounter{equation}{3}
\begin{align}\label{equ:transform2}
\mathbf C_b^t &= {\mathbf T_3}\left( \gamma  \right){\mathbf T_2}\left( \beta \right){\mathbf T_1}\left(\alpha\right)=\left[{\begin{array}{*{20}{c}}
{\cos \alpha \cos \beta }&{  \sin \alpha \cos \beta }&{-\sin \beta }\\
{\cos \alpha \sin \beta \sin \gamma -\sin \alpha \cos \gamma  }&{\sin \alpha\sin \beta \sin \gamma  + \cos \alpha \cos \gamma }&{\cos \beta \sin \gamma }\\
{\sin \alpha \sin\gamma  + \cos \alpha \sin \beta \cos \gamma }&{\sin \alpha \sin \beta \cos\gamma-\cos \alpha \sin \gamma}&{  \cos \beta \cos \gamma }
\end{array}}\right].
\end{align}
\setcounter{equation}{\value{MYtempeqncnt}}
\addtocounter{equation}{1}
\hrulefill
\vspace*{4pt}
\end{figure*}
\begin{figure*}[!t]
\normalsize
\setcounter{MYtempeqncnt}{\value{equation}}
\setcounter{equation}{9}
\begin{align}\label{equ:Isolation1}
{\boldsymbol\omega _{ut}}=&\mathbf C_b^t\boldsymbol \omega _{ub}=  {\mathbf T_3}\left( \gamma  \right){\mathbf T_2}\left( \beta \right){\mathbf T_1}\left(\alpha\right){\boldsymbol\omega _{ub}}\notag\\
=&\left[{\begin{array}{*{20}{c}}
1&0&0\\
0&{\cos \gamma }&{\sin \gamma }\\
0&{ - \sin \gamma }&{\cos \gamma }
\end{array}} \right]\left[{\begin{array}{*{20}{c}}
{\cos \beta }&0&{ - \sin \beta }\\
0&1&0\\
{\sin \beta }&0&{\cos \beta }
\end{array}} \right]\left[ {\begin{array}{*{20}{c}}
{\cos \alpha }&{\sin \alpha }&0\\
{ - \sin \alpha }&{\cos \alpha }&0\\
0&0&1
\end{array}} \right]\left[ \begin{array}{l}
{\omega _{ubx}}\\
{\omega _{uby}}\\
{\omega _{ubz}}
\end{array} \right].
\end{align}
\setcounter{equation}{\value{MYtempeqncnt}}
\addtocounter{equation}{1}
\hrulefill
\vspace*{4pt}
\end{figure*}

According to the position relationships of UAV and satellite, we can obtain the Euler angles $o,\ e,\ v $ of the spatial beam relative to the $n$-frame as
\begin{align}\label{equ:Alignment1}\setcounter{equation}{4}
\left\{ \begin{array}{l}
o = {180^ \circ } + \tan^{ - 1}\left( {\frac{{\tan (\lambda  - {\lambda _s})}}{{\sin \vartheta }}} \right),\\
e = {\tan ^{ - 1}}\left( {\frac{{\cos \varphi \cos (\lambda  - {\lambda _s}) - \frac{{{R_E}}}{{{R_E} + {h_E}}}}}{{\sqrt {1 - {{\left[ {\cos \varphi \cos (\lambda  - {\lambda _s})} \right]}^2}} }}} \right),\\
v = {\tan ^{ - 1}}\left( {\frac{{\sin (\lambda  - {\lambda _s})}}{{\tan \vartheta }}} \right),
\end{array} \right.
\end{align}
where  $\vartheta$ is the UAV latitude, $\lambda$ is UAV longitude, $\lambda_s$ is longitude of target satellite, $R_E$ and $h_E$ are the earth radius and the UAV altitude.

Similarly, we can derive the coordinates transformation matrix between the $n$-frame and the $t$-frame $\mathbf C_n^t$ as
\begin{align}\label{equ:transform3}
\mathbf C_n^t &= {\mathbf T_3}\left( v  \right){\mathbf T_2}\left( e  \right){\mathbf T_1}\left(o \right).
\end{align}

Furthermore, when the UAV attitude information $\left( {\psi ,\theta ,\phi } \right)$ is obtained (the derivation is presented in Appendix B),  the coordinates transformation matrix between the $n$-frame and the $b$-frame $\mathbf C_n^b$ can be expressed as
\begin{align}\label{equ:transform1}
\mathbf C_n^b &= {\mathbf T_3}\left( \phi  \right){\mathbf T_2}\left( \theta  \right){\mathbf T_1}\left( \psi  \right).
\end{align}

As shown in  Fig. \ref{fig:frame_relaltionship}, the coordinates transformation from the $n$-frame to the $t$-frame can be decomposed into the transformation from $n$-frame to the $b$-frame and the transformation from the $b$-frame to the $t$-frame, namely,
\begin{align}\label{equ:transform4}
\mathbf C_n^t = \mathbf C_n^b\mathbf C_b^t,\quad \text{or} \quad \mathbf C_b^t = (\mathbf C_n^b)^{-1}\mathbf C_n^t.
\end{align}

From (\ref{equ:transform4}), the final angle in the $b$-frame for aligning the spatial beam of UAV to the target satellite can be derived as
\begin{align}\label{equ:transform5}
\left\{ {\begin{array}{*{20}{l}}
{\alpha  = \arctan  \frac{T_{12}}{T_{11}}},\\
{\beta  =  - \arcsin {T_{13}}},\\
{\gamma  = \arctan \frac{T_{23}}{T_{33}}},
\end{array}} \right.
\end{align}
where $T_{m,n}$ is the $(m,n)$-th element in $\mathbf C_b^t$.
\subsection{Dynamic Isolation}\label{sec:Attitude}

When UAV navigates, its attitude will change and would affect beam pointing. The corresponding angular rate  in the $b$-frame, denoted as ${\boldsymbol\omega _{ub}}$, is coupled with the direction of the spatial beam through geometric constraints and friction constraints, and could be measured by gyros. Let us further denote the coupled angular rate of the spatial beam due to UAV navigation in the $t$-frame as ${\boldsymbol\omega _{ut}}$. Then, it can be derived as \eqref{equ:Isolation1}, shown on the top of the page. Moreover, in \eqref{equ:Isolation1}, $\omega _{ubx}$, $\omega _{uby}$ and $\omega _{ubz}$ are the elements of ${\boldsymbol\omega _{ub}}$.

\begin{figure*}[t]
\centering
\includegraphics[width=140mm]{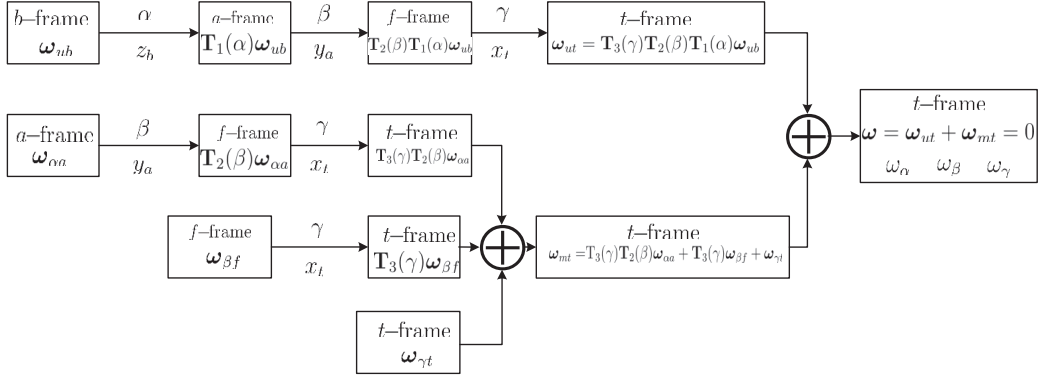}
\caption{The procedure of dynamic isolation.}
\label{fig:isolation}
\end{figure*}

Next, we propose the dynamic isolation algorithm to compensate ${\boldsymbol\omega _{ut}}$ as well as release the effects of UAV navigation, and the detailed scheme is shown in Fig. \ref{fig:isolation}. Specially, the control monitors of the azimuth angle, the polarization angle, and the elevation angle are jointly utilized to make the spatial beam point to the target satellite.

Denote $\boldsymbol\omega _{\alpha a}$, $\boldsymbol\omega _{\beta f}$ and $\boldsymbol\omega _{\gamma t}$ as the control monitor outputs of the azimuth angle, the polarization angle, and the elevation angle respectively. According to the inter-relations of the $a$-frame,  the $f$-frame, and the $t$-frame, the azimuth angular rate in the $t$-frame, denoted as ${\boldsymbol\omega _{\alpha t}}$,   can be obtained from azimuth angle control monitor as
\begin{align}\label{equ:Isolation4}
\setcounter{equation}{10}
\!\!\!&{\boldsymbol\omega _{\alpha t}}={\mathbf T_3}\left( \gamma  \right){\mathbf T_2}\left( \beta \right)\boldsymbol\omega _{\alpha a}\notag \\
&\!\!\!=\!\!\!\left[ {\begin{array}{*{20}{c}}
1&0&0\\
0&{\cos \gamma }&{\sin \gamma }\\
0&{ - \sin \gamma }&{\cos \gamma }
\end{array}} \right]\!\!\!\left[ {\begin{array}{*{20}{c}}
{\cos \beta }&0&{ - \sin \beta }\\
0&1&0\\
{\sin \beta }&0&{\cos \beta }
\end{array}} \right]\!\!\!\left[\!\!\!\begin{array}{c}
0\\
0\\
{\omega _\alpha }
\end{array}\!\!\!\right],
\end{align}
while the elevation angular rate in the $t$-frame,  denoted as ${\boldsymbol\omega _{\beta t}}$, can be obtained  from the elevation control monitor as
\begin{align}\label{equ:Isolation7}
{\boldsymbol\omega _{\beta t}}={\mathbf T_3}\left( \gamma  \right)\boldsymbol\omega _{\beta f}=\left[{\begin{array}{*{20}{c}}
1&0&0\\
0&{\cos \gamma }&{\sin \gamma }\\
0&{ - \sin \gamma }&{\cos \gamma }
\end{array}} \right]\left[\!\!\!\begin{array}{c}
0\\
{\omega _\beta }\\
0
\end{array} \!\!\!\right].
\end{align}
Moreover,  the polarization angular rate in the $t$-frame can be directly obtained by ${\boldsymbol\omega _{\gamma t}}=\left[\omega_\gamma\ 0\ 0 \right]^T$. 

Then, the total beam pointing angular rates from the control monitors in the $t$-frame, denoted as ${\boldsymbol\omega _{mt}}$, are the sum of ${\boldsymbol\omega _{\alpha t}}$, $\boldsymbol\omega _{jt\beta}$, and ${\boldsymbol\omega _{\gamma t}}$, i.e.,
\begin{align}\label{equ:Isolation6}
{\boldsymbol\omega _{mt}}={\boldsymbol\omega _{\gamma t}}+{\mathbf T_3}\left( \gamma  \right)\boldsymbol\omega _{\beta f}+{\mathbf T_3}\left( \gamma  \right){\mathbf T_2}\left( \beta \right)\boldsymbol\omega _{\alpha a}.
\end{align}

Take the $t$-frame as the reference, the eventual physical angular rate of beam pointing, denoted as $\boldsymbol \omega$, is the result of a combination of  the UAV's own navigation and the control monitor's outputs,  namely,
\begin{align}\label{equ:Isolation5}
&\boldsymbol \omega={\boldsymbol\omega _{mt}}+{\boldsymbol\omega _{ut}}.
\end{align}

To overcome the effect of UAV attitude variation, it is required that  $\boldsymbol \omega=0$ such that the spatial beam is always pointing to the target satellite. After some tedious computations, the angular rates of the control monitors can be obtained from~\eqref{equ:Isolation5} as
\begin{align}\label{equ:Isolation8}
\!\!\!\left\{ {\begin{array}{*{20}{l}}
{{\omega _\alpha } \!\!= \!\! - \cos \alpha  \cdot \tan\beta  \cdot {\omega _{ubx}}\! - \!\sin \alpha  \cdot \tan\beta  \cdot {\omega _{uby}}\! - \! {\omega _{ubz}}},\\
{{\omega _\beta } = \sin \alpha  \cdot {\omega _{ubx}} - \cos \alpha  \cdot {\omega _{uby}}},\\
{{\omega _\gamma } = ( - \cos \alpha  \cdot {\omega _{ubx}} - \sin \alpha  \cdot {\omega _{uby}})\sec \beta }.
\end{array}} \right.
\end{align}
\begin{remark}
The mechanical adjustment is performed according to the attitude information rather than the training symbols, which is a blind way for beam tracking.
\end{remark}
\section {Fine Beam Alignment with Electrical Adjustment}\label{sec:simultaneous}
Due to the system error and the measurement error,  purely counting on mechanical adjustment  is not
sufficient to yield an accurate beam pointing. To further improve the precision of beam pointing, we propose an electrical adjustment method.
\subsection{ Electrical Adjustment with Simultaneous Perturbation}

As mentioned earlier, we consider the hybrid scheme to decrease the system cost and power consumption. Therefore, the received signal at UAV  can be expressed as
 \begin{align}\label{equ:received}
 y=\mathbf w^H \mathbf h s+\mathbf w^H\mathbf n,
\end{align}
where $\mathbf h=\textup{vec} (\mathbf H)$, $\mathbf w$ is the $MN\times 1$ receive beamforming vector  with $|\mathbf w_{i}|=1$, $s$ is the transmitted signal of the satellite, and $\mathbf n$ is the $M N\times1$ noise vector.

Then, the instant power of the received signal at UAV can be expressed as
\begin{align}\label{equ:power}
P=|y|^2=\mathbf w^H \mathbf R_s \mathbf w+\mathbf w^H \mathbf R_n \mathbf w,
\end{align} 
where  $\mathbf R_s=\mathbf h ss\mathbf h^H $ and $\mathbf R_n=\mathbf n\mathbf n^H$ are the temporal covariance matrices of the signal and the noise respectively.

Let $[\alpha,\beta]$ be the relative location of the target satellite after mechanical adjustment. Since mechanical adjustment could approximately stabilize the beam direction near the normal of UPA, the value of $\alpha$ and $\beta$ would be small.
To achieve a superior quality of service in UAV-satellite communication, the power of the received signal at UAV should be made as high as possible to guarantee the beam pointing in a precise direction \cite{beamalignment}.

It has been demonstrated in \cite{xie} that $P$ can be maximized by aligning the beam direction  with the desired signal source direction,   i.e., $\mathbf w=\textup{vec}\left(\mathbf A(\alpha, \beta)\right)$. 
In this scheme, the phase of the $(m,n)$-th element of $\mathbf A(\alpha, \beta)$, denoted as $p_{(m,n)}$ can be expressed as
\begin{align}\label{equ:phase}
p_{(m,n)}&=\frac{2\pi d\sin \alpha}{\lambda_c}\left[(m-1)\cos \beta+(n-1)\sin\beta\right]\notag\\
&=\frac{2\pi d\sin \alpha}{\lambda_c}\sqrt{(m-1)^2+(n-1)^2}\notag\\
&.\sin\left[\beta+\arcsin \frac{m-1}{\sqrt{(m-1)^2+(n-1)^2}}\right].
\end{align}

\begin{figure*}[!t]
\normalsize
\setcounter{MYtempeqncnt}{\value{equation}}
\setcounter{equation}{18}

\begin{align}\label{equ:gradient2}
\hat g_t(\mathbf  w_{k})=\frac{P(\mathbf  w_{k}+\frac{b\mathbf D \xi_k}{(k+1)^\Omega}+\frac{c\boldsymbol \Delta_k}{(k+1)^\Omega}/)-P(\mathbf  w_{k}-\frac{b\mathbf D \xi_k}{(k+1)^\Omega}-\frac{c\boldsymbol \Delta_k}{(k+1)^\Omega}/)}{2\left[\frac{b\mathbf D \xi_k}{(k+1)^\Omega}+\frac{c\boldsymbol \Delta_k}{(k+1)^\Omega}\right]}.
\end{align}
\setcounter{equation}{\value{MYtempeqncnt}}
\addtocounter{equation}{1}
\hrulefill
\vspace*{4pt}
\end{figure*}
From \eqref{equ:phase}, the DOA information is required for receive beamforming. However, the single RF chain at UAV restricts the application of many advanced DOA estimation algorithms, such as MUSIC, ESPRIT, and Capon \cite{2D-ESPRIT,trackingangle1,trackingangle2,gao4}. For single RF chain receiver, a zero-knowledge beamforming method was suggested in \cite{sim}, but this method is time-consuming due to sequential perturbation of the phase shifters. Besides, the convergence speed would become much slower with the increase of the massive array antennas. A simultaneous perturbation method  was proposed in \cite{simul} to accelerate the convergence speed. However, the method is fully isotropic and still requires many iterations for convergence. Here, we propose to utilize the array structure to further accelerate the converge speed for UAV with massive array antennas.

Note from \eqref{equ:phase} that the values of the phase shifters for beam alignment are related to the antenna locations $(m,n)$. Hence, we could utilize the array antenna structure to accelerate the convergence speed. Specially, the instant power of the received signals are utilized for the electrical adjustment instead of the training symbols, which is also a blind way.

Denote $\hat g_k(\mathbf  w_{k})$ as the estimation of the gradient $\partial P(\mathbf  w_{k})/\partial \mathbf  w_{k}$, which can be derived from two noisy measurements of the received signal power as \eqref{equ:gradient2}, shown on the top of the page.

Moreover, in \eqref{equ:gradient2}, $\xi_k$ and the elements of $\boldsymbol \Delta_k$ are  random perturbation parameter at the $k$-th iteration following a Bernoulli $\pm 1$ distribution with probability $1/2$. Meanwhile, $b$, $c$ as well as $\Omega$ are all the small perturbation parameters, and $\mathbf D$ can be expressed as
\begin{align}\label{equ:antenna_location}
\mathbf D=\begin{bmatrix}
d_{11} & \cdots & d_{1N} \\
\vdots & \ddots & \vdots \\
d_{M1}&\cdots &d_{MN}
\end{bmatrix},
\end{align}
where $d_{mn}=\sqrt{(m -1)^2+ (n-1)^2}$.

Then, the received beamforming vector can be adjusted according to the following formula
\begin{align}\label{equ:gradient1}
\mathbf  w_{k+1}=\mathbf  w_{k}-\eta_k\hat g_k(\mathbf  w_{k}),
\end{align}
where $\eta_k=a/(\zeta+k)^\xi$ is the step size for the adjustment at the $k$-th iteration, and $\xi$, $\zeta$ are step perturbation parameters.
The perturbation parameters in \eqref{equ:gradient2} and \eqref{equ:gradient1} can be obtained by the guide lines in \cite{rules}.

The detailed steps of the proposed method are shown in Algorithm \ref{alg:algorithm2}. Since the proposed electrical adjustment utilizes the array structure, it can be named as \emph{array structure based simultaneous perturbation} (ASSP).

\begin{remark}
Since the value of $\alpha$ and $\beta$ are small, the time for convergence would  not be large. Besides, by utilizing the array structure, the convergence speed of the proposed algorithm is  not relevant to the number of the phase shifter or equivalently the number of the array antennas, and would be much faster than the traditional techniques in~\cite{sim,simul}.
\end{remark}
\subsection{ Complex Gains Estimation}
Through the joint mechanical adjustment and electrical adjustment, the spatial beam can be  pointed to the target satellite without training, i.e., a blind way.  With the tracked beam, we can further derive the gain information by sending training symbols  towards the target satellite as did in \cite{xie}. Since there is only one dominant beam for transmission, the  minimum size of the training sequence is $1$ for both uplink and downlink channel. 

\begin{algorithm}[t]
\caption{: Electrical adjustment}
\label{alg:algorithm2}
\begin{itemize}
\item \textbf{Step 1:} Initialization: initialize the spatial beam direction of the array antenna as $\mathbf w =\mathbf 1_{M N\times1}$, and obtain the UAV latitude $\vartheta$, the UAV longitude $\lambda$ and the satellite longitude $\lambda_s$;
\item \textbf{Step 2:} Beam coarse alignment: adopt beam stabilization \eqref{equ:transform5} and dynamic isolation \eqref{equ:Isolation7}   to realize the rough alignment of the spatial beam;
\item \textbf{Step 3:} Beam fine alignment: perform the array structure aided DOA tracking. Set $k=0$, and pick parameters $a,\ b,\ c,\ \zeta,\ \Omega,\ \xi$ according to the guide lines \cite{rules};
\item \textbf{Step 4:} Simultaneous perturbation: generate the perturbation scalar $\xi_k$ and  $\boldsymbol \Delta_k$, and derive two instant power measurements of the UAV received signals $P(\mathbf w^+)$ and $P(\mathbf w^-)$, where $\mathbf w^+=\mathbf  w_{k}+b/(k+1)^\Omega\mathbf D \xi_k+c/(k+1)^\Omega\boldsymbol \Delta_k$  and $\mathbf w^-=\mathbf  w_{k}-b/(k+1)^\Omega\mathbf D \xi_k-c/(k+1)^\Omega\boldsymbol \Delta_k$;
\item \textbf{Step 5:} The beam pointing updating: estimate the gradient $\hat g_t(\mathbf  w_{k})$ using \eqref{equ:gradient2} and update
the beam direction $\mathbf w_{k+1}$ by \eqref{equ:gradient1};
\item \textbf{Step 6:} Decision: terminate algorithm when the power of the received signal increases little, or else return to Step~3.
\end{itemize}
\end{algorithm}

\section{Simulations and Results} \label{sec:simulation}
In this section, numerical results are provided to demonstrate the effectiveness of the proposed method. We set $M=128$,  $N=64$ and $d=\frac{\lambda}{2}$, and the channel vector between UAV and target satellite can be produced according to (\ref{equ:channelmodel}). The experimental platform, which consists of the outdoor part and the indoor part, is installed at the three-axis flight simulation turntable to imitate the navigation of UAV. The outdoor part are composed of XW-ADU3601, XW-IMU5220 and XW-ADU7612 \cite{datasheet1,datasheet2,datasheet3}. Specifically,
XW-ADU3601 is a two-antenna GPS outputting the measurement of $\psi_m$; XW-IMU5220 is an MEMS inertial measurement unit that can output the attitude angular rates $\boldsymbol\omega_{ub}$ and the measurements of $\theta_m$ and $\phi_m$; XW-ADU7612 is an attitude and heading reference system, which can provide the attitude to an accuracy of $0.1^\circ$ as comparison. The DSP processor runs in the indoor parts with the frequency 400 MHz, which is ACU of the platform. Note that the platform can provide measurements of the attitude information, and it can be used to verify the validity of the mechanical adjustment. The target satellite is set as AsiaSat-3S \cite{datasheet4}  with longitude $105.5^\circ$, while the testing location is Xi'an shaanxi with latitude $34.27^\circ$ and longitude $108.95^\circ$. According to \eqref{equ:Alignment1}, the Euler angles of the beam alignment is $a=6.11^\circ$, $e=50.1^\circ$, and $v=-5.05^\circ$.

\begin{figure}[t]
\centering
\includegraphics[height=65mm,width=90mm]{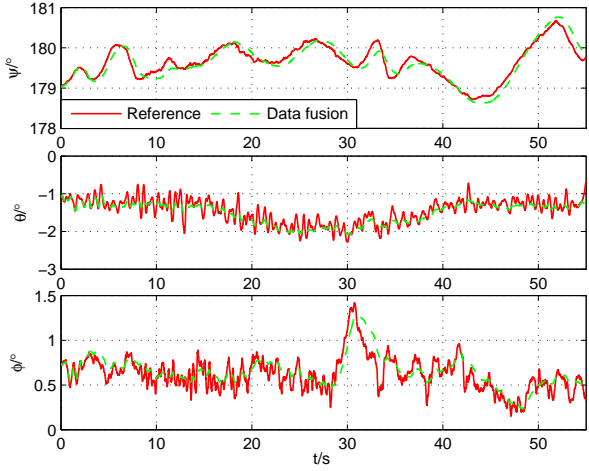}
\caption{The performance of attitude determination.}
\label{fig:1}
\end{figure}

\begin{figure}[t]
\centering
\includegraphics[height=65mm,width=90mm]{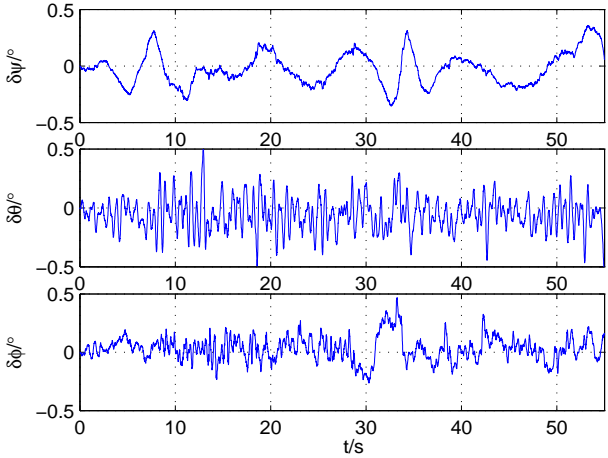}
\caption{The error of the estimated attitude.}
\label{fig:2}
\end{figure}

\begin{figure}[t]
\centering
\includegraphics[height=65mm,width=90mm]{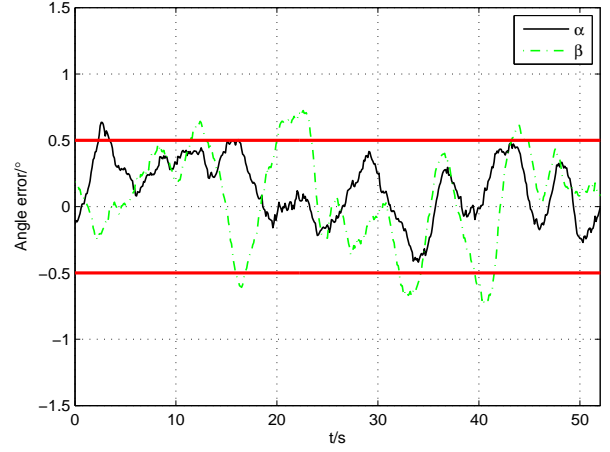}
\caption{The errors of the beam pointing from mechanical adjustment.}
\label{fig:error_analysis}
\end{figure}

Fig. \ref{fig:1} shows the attitude from Kalman filter based data fusion, where the attitude from the XW-ADU7612 is also displayed as comparison, while Fig. \ref{fig:2} shows the corresponding attitude error. It can be seen from Fig. \ref{fig:1} that the regulation of attitude variation from data fusion coincides with the practical attitude value, which verifies that the data fusion can effectively integrate the measurements from different sensors and provide relatively precise attitude information. However, we can see clearly from Fig. \ref{fig:2} that there are still attitude errors from data fusion, and the maximum error of the attitude reaches  $0.5^\circ$, which is consistent with the result in \eqref{equ:attitude16}-\eqref{equ:attitude18}.

With the tracked attitude, we then derive the beam pointing through mechanical adjustment, and the beam pointing errors of the azimuth angle $\alpha$ and the elevation angle $\beta$ are shown in Fig.~\ref{fig:error_analysis}. It can be seen that the pointing errors of both $\alpha$ and $\beta$ are within $0.5^\circ$ in most time, and mechanical adjustment can only approximately make the beam point to the target satellite.  The reason is that the attitude error from data fusion cannot be neglected, which would be coupled into beam pointing and make the beam deviate from the target satellite.

\begin{figure}[t]
\centering
\includegraphics[height=65mm,width=90mm]{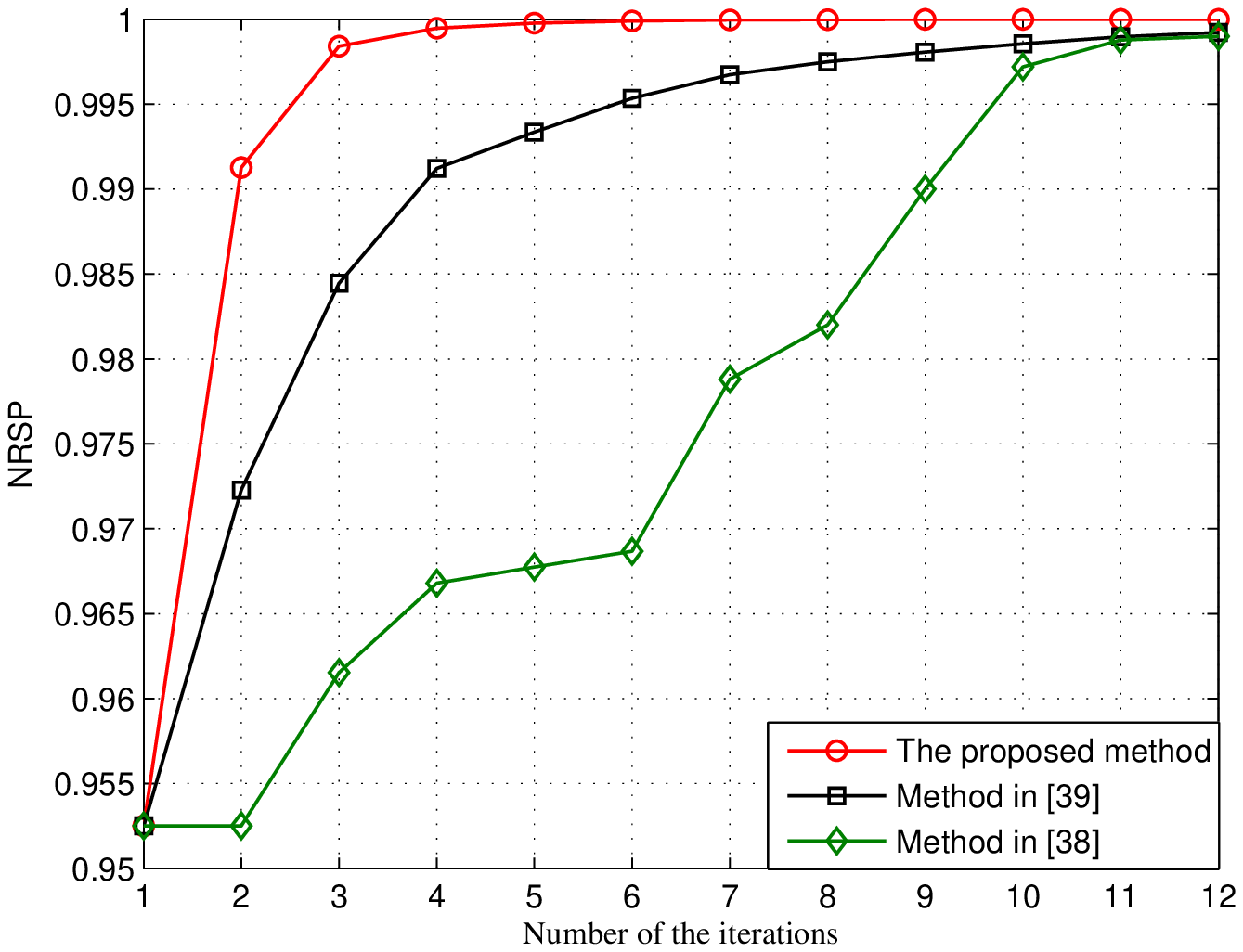}
\caption{The performance of the electrical adjustment, where SNR$=20$dB.}
\label{fig:111}
\end{figure}

\begin{figure*}[!t]
\normalsize
\setcounter{MYtempeqncnt}{\value{equation}}
\setcounter{equation}{21}
\begin{align}\label{equ:attitude1}
\left[\begin{array}{*{20}{c}} \omega _{ubx}(\tau)\\\omega _{uby}(\tau)\\\omega _{ubz}(\tau)\end{array}\right] &=\left[ \begin{array}{*{20}{c}}
{\dot \phi(\tau) }\\0\\0\end{array} \right] + {\mathbf T_3}\left[ \phi(\tau-1)  \right]\left[ {\begin{array}{*{20}{c}}0\\\dot \theta(\tau)\\0\end{array}} \right] +{\mathbf T_3}\left[\phi(\tau-1)  \right]{\mathbf T_2}\left[ \theta (\tau-1) \right]\left[ {\begin{array}{*{20}{c}}0\\0\\{\dot \psi(\tau)}
\end{array}} \right]\notag\\&= \left[ {\begin{array}{*{20}{c}}
1&{\sin \phi (\tau-1)\tan \theta(\tau-1) }&{\cos \phi(\tau-1) \tan \theta (\tau-1)}\\
0&{\cos \phi (\tau-1) }&{ - \sin \phi (\tau-1)}\\
0&{\sin \phi(\tau-1) /\cos \theta(\tau-1) }&{\cos \phi(\tau-1) /\cos \theta(\tau-1)}
\end{array}} \right]\left[ {\begin{array}{*{20}{c}}
{{\omega _{ubx}(\tau)}}\\
{{\omega _{uby}(\tau)}}\\
{{\omega _{ubz}(\tau)}}
\end{array}} \right].
\end{align}
\setcounter{equation}{\value{MYtempeqncnt}}
\addtocounter{equation}{1}
\hrulefill
\vspace*{4pt}
\end{figure*}

\begin{figure}[t]
\centering
\includegraphics[height=65mm,width=90mm]{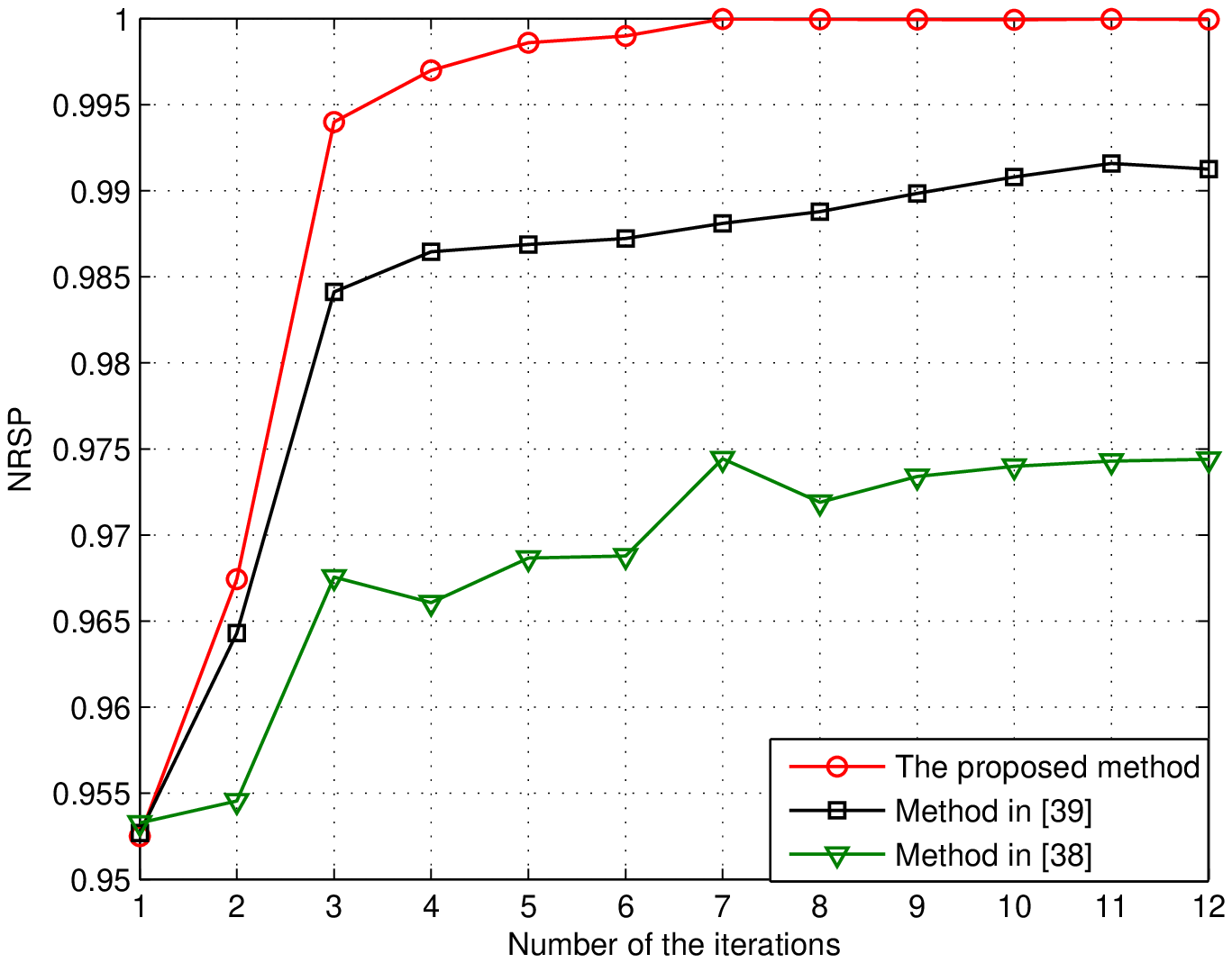}
\caption{The performance of the electrical adjustment, where SNR$=$10dB.}
\label{fig:snr}
\end{figure}
\begin{figure}[t]
\centering
\includegraphics[height=65mm,width=90mm]{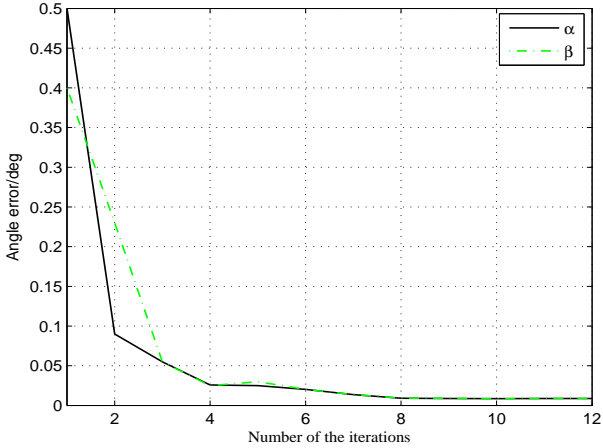}
\caption{The angle error of the electrical adjustment, where SNR$=$10dB.}
\label{fig:ea}
\end{figure}
Next, we show the performance of electrical adjustment in Fig. \ref{fig:111} and Fig.~\ref{fig:snr} with SNR$=$10 dB and SNR$=$20 dB respectively, where the method in \cite{sim} and \cite{simul} are also displayed as comparison. According to the guide lines in~\cite{rules}, the step parameters are set as $a=0.7,b=0.02,c=0.01,\zeta=0.1,\tau=0.1,\xi=0.602$.  The performance metric is the normalized received signal power (NRSP).  The prior NRSP is 0.952, which further verifies that the mechanical adjustment could approximately point the spatial beam to the target satellite.  It can be seen that all the perturbation methods can converge to the maximum received power, i.e., the spatial beam of UAV is pointed to the practical DOA of the incident signals. The proposed method can converge with much faster speed than other two methods, since the array structure is utilized for beam tracking. Besides, we can find that the number of iteration is increasing with the decrease of SNR for all the displayed methods. The number of iteration for convergence is 4 for SNR$=$20dB, while it increases to 7 at SNR$=$10 dB for the proposed method. However, the convergence of the methods in \cite{sim} and \cite{simul} is much worse compared to their higher SNR case. This is because for low SNR case, the noise is dominant and would affect the convergence speed of the electrical adjustment.

In the last example, we show the final angle error of the proposed simultaneous perturbation with SNR$=$10 dB in Fig.~\ref{fig:ea}. We see that
with the increase of the iteration number, the angle errors of both $\alpha$ and $\beta$ become smaller, and the eventual angle errors stay in the magnitude of $10^{-2}$, which is consistent with the result in Fig. \ref{fig:111}. Hence, through the joint mechanical and electrical adjustment, the spatial beam can be precisely pointed to the target satellite.

\section{Conclusions} \label{sec:conclusions}
In this paper, we proposed a blind beam tracking method for Ka-band UAV-satellite communication system with hybrid massive antennas. The proposed method mainly consists of two procedures: mechanical adjustment and electrical adjustment. The former one, including the beam stabilization and dynamic isolation, is used to overcome the effect of UAV navigation and realize the coarse alignment, while the latter one is used to further improve the tracking precision and realize the fine alignment. Moreover, tracking the spatial beam can be simplified to tracking the DOA information of the satellite, which greatly decreases the training overhead and improves the system efficiency. Various  examples are provided to verify the efficiency of the proposed method.

\appendices
\section{ Attitude  Determination Principles of The Low-Cost Sensors}
(1) {MEMS Gyros:}
The three MEMS gyros (roll gyro, pitch gyro and yaw gyro) are installed near the UAVCG in the three directions of the $b$-frame, which can measure the attitude variation of UAV. Denote the angular rate outputs of the gyros as $\boldsymbol \omega _{ub}=\left[\omega _{ubx}(\tau)\ \omega _{uby}(\tau)\ \omega _{ubz}(\tau)\right]^T$, and  denote the variation of the UAV attitudes as $\left[\dot{\phi}(\tau)\ \dot{\theta}(\tau) \ \dot{\psi}(\tau)\right]$. Denote $\tau$ as the time index. According to the definitions of the reference frames, the relationship between $\boldsymbol \omega _{ub}$ and $\left[\dot{\phi}(\tau)\ \dot{\theta}(\tau) \ \dot{\psi}(\tau)\right]$ can be expressed as \eqref{equ:attitude1}, shown on the top of the page.


Then, we can obtain  a rough measurement of the three-dimension attitude angles, denoted as $\phi_m(\tau)$, $\theta_m(\tau)$, and $ \psi_m(\tau)$, from the gyro integration as
\begin{align}
\left[ {\begin{array}{*{20}{c}}
{ \phi_m(\tau)} \\
{ \theta_m(\tau) }\\
{ \psi_m(\tau) }
\end{array}} \right] = \left[ {\begin{array}{*{20}{c}}
{ \phi_m(\tau-1) }\\
{ \theta _m(\tau-1)}\\
{ \psi _m(\tau-1)}
\end{array}} \right]+\left[ {\begin{array}{*{20}{c}}
{\dot \phi (\tau)}\\
{\dot \theta (\tau)}\\
{\dot \psi(\tau) }
\end{array}} \right].
\end{align}
\begin{figure*}[!t]
\normalsize
\setcounter{MYtempeqncnt}{\value{equation}}
\setcounter{equation}{30}
\begin{align}\label{equ:attitude10}
{\boldsymbol\Xi}[\mathbf q(\tau-1)] = \left[ {\begin{array}{*{20}{c}}
{{q_4(\tau-1)}{{\bf{I}}_{3 \times 3}} + \left[ {\begin{array}{*{20}{c}}
0&{ - {q_3(\tau-1)}}&{{q_2(\tau-1)}}\\
{{q_3(\tau-1)}}&0&{ - {q_1(\tau-1)}}\\
{ - {q_2(\tau-1)}}&{{q_1(\tau-1)}}&0
\end{array}} \right]}\\
{ - {{\boldsymbol{\rho }}^{\rm{T}}}}
\end{array}} \right].
\end{align}
\setcounter{equation}{\value{MYtempeqncnt}}
\addtocounter{equation}{1}
\hrulefill
\vspace*{4pt}
\end{figure*}

(2) {Accelerator:} The accelerator is also installed at UAVCG, which  can derive the attitude of the UAV by measuring the local gravity acceleration directly. Denote $\mathbf f(\tau)=\left[ f_x(\tau),f_y(\tau),f_z(\tau) \right]^T$ as the measurement of the accelerator, and the relationship between the output of accelerator and the attitude of UAV can be expressed as
\begin{align}\label{equ:attitude3}
\setcounter{equation}{23}
\mathbf f(\tau)=\left[ {\begin{array}{*{20}{c}}
f_x(\tau)\\f_y(\tau)\\f_z(\tau)\end{array}} \right] \approx  - g\left[ {\begin{array}{*{20}{c}}
{ - \sin \theta (\tau)}\\{\sin \phi(\tau) \cos \theta(\tau) }\\{\cos \phi(\tau) \cos \theta(\tau) }\end{array}} \right],
\end{align}
where $g$ is the gravitational acceleration.

Then, the measured pitch angle and the roll angle can be obtained by
\begin{align}\label{equ:attitude4}
&{\theta_m(\tau) } = \arcsin \left[\frac{{{f_x(\tau)}}}{g}\right],\\
&{\phi_m (\tau)} = \arctan \left[\frac{{{f_y(\tau)}}}{{{f_z(\tau)}}}\right].
\end{align}

(3) {GPS:}
When two GPS antennas are installed at the front and back of UAV, they can provide measured yaw angle by using carrier-phase differential technology. When the baseline length between the two GPS antennas is $d$, the coordinate of baseline can be expressed as $[d \ 0\ 0]$ in the $b$-frame. Meanwhile, we can derive the coordinate of baseline in the $n$-frame as $[x(\tau) \ y(\tau)\ z(\tau)]$.
Then, the coordinates of the same baseline in the $b$-frame and the $n$-frame can be transformed by $\mathbf C_n^b$ as
\begin{align}\label{equ:attitude5}
\left[ {\begin{array}{*{20}{c}}
d\\
0\\
0
\end{array}} \right] =\mathbf C_n^b \left[ {\begin{array}{*{20}{c}}
x(\tau)\\
y(\tau)\\
z(\tau)
\end{array}} \right].
\end{align}

Then, we can acquire the measured yaw angle from \eqref{equ:attitude5} as
\begin{align}\label{equ:attitude6}
\psi_m (\tau)  =  - \arctan \left[\frac{z(\tau)}{x(\tau)}\right].
\end{align}

\section{Low Cost Attitude Determination with Data Fusion }

The low cost attitude sensors can measure the attitude information. However, there would exist large error due to the imprecision of  the low-cost gyros, e.g., the drift error and sideslip angle. Therefore, we try to utilize the Kalman filter based data fusion to derive an accurate UAV attitude.

Direction cosine matrix, quaternion and Euler angle are the most commonly used methods for attitude description, among which the quaternion method not only has low computation complexity, but can  also be realtimely updated with high accuracy. Hence, we adopt the quaternion for attitude representation.

The quaternion uses four parameters, denoted as $q_1$, $q_2$, $q_3$, and $q_4$,  to represent the three dimension attitude of the UAV, which is defined as
\begin{align}\label{equ:attitude71}
\mathbf q = {\left[ {\begin{array}{*{20}{c}}q_1&\boldsymbol{\rho}\end{array}} \right]^{\rm{T}}},
\end{align}
where $\boldsymbol{\rho} = {\left[ {\begin{array}{*{20}{c}}q_2&q_3&{q_4}\end{array}} \right]^{\rm{T}}}$.

With the quaternion, the attitude updating equation can be represented by
\begin{align}\label{equ:attitude9}
\mathbf{q}(\tau) = \mathbf{q}(\tau-1)+\frac{T_s}{2}{\boldsymbol\Xi   ^{\rm{T}}}[{\bf{q}}(\tau-1)]{\boldsymbol{\omega_{ub} }}(\tau),
\end{align}
where $T_s$ is the sampling interval. Moreover, ${\boldsymbol\Xi}[\mathbf q(\tau-1)]$ can be derived as \eqref{equ:attitude10}, shown on the top of the page.

After some tedious transformations of \eqref{equ:attitude9}, we can further derive that
\begin{align}\label{equ:attitude13}
\setcounter{equation}{31}
\mathbf{q}(\tau)=\boldsymbol \Gamma\mathbf{q}(\tau-1) =\mathbf{q}(\tau-1),
\end{align}
where

$\Gamma=\left( \mathbf{I}_{4\times 4}+\frac{T_s}{2}\left[ {\begin{array}{*{20}{c}}
0&{ - {\omega _{ub x}}}&{ - {\omega _{ub y}}}&{ - {\omega _{ub z}}}\\
{{\omega _{ub x}}}&0&{{\omega _{ub z}}}&{ - {\omega _{ub y}}}\\
{{\omega _{ub y}}}&{ - {\omega _{ub z}}}&0&{{\omega _{ub x}}}\\
{{\omega _{ub z}}}&{{\omega _{ub y}}}&{ - {\omega _{ub x}}}&0
\end{array}} \right]\right)$.

Therefore, the \emph{system equation} can be expressed as
\begin{align}\label{equ:systemmodel}
\mathbf{q}(\tau) =\boldsymbol \Gamma\mathbf{q}(\tau-1)+\boldsymbol \chi,
\end{align}
where $\boldsymbol \chi$ is the system noise vector, and each element belongs to  $\mathcal {CN}(0,\mathbf{Q}_{\boldsymbol \chi})$.

When the quaternion is used for attitude updating, it should meet the normalized condition:
\begin{align}\label{equ:attitude7}
{q_1}^2 + {q_2}^2 + {q_3}^2 + {q_4}^2 = 1.
\end{align}
Then, the normalized quaternion can be expressed as
\begin{align}\label{equ:attitude8}
\mathbf q' = \frac{\mathbf q}{\sqrt{\mathbf q^T \mathbf q}}.
\end{align}

Since we can directly obtain coarse values of the attitude from the low-cost sensors, the measurement of the quaternion can be obtained from
\begin{align}\label{equ:attitude15}
&q_{0 m} ={{\rm{cos\frac{\psi_m }{2}cos\frac{\theta_m }{2}cos\frac{\phi_m }{2}+ sin\frac{\psi_m }{2}sin\frac{\theta_m }{2}sin\frac{\phi_m }{2}}}},\\
&q_{1 m} ={{\rm{cos\frac{\psi_m }{2}cos\frac{\theta_m }{2}sin\frac{\phi_m }{2} - sin\frac{\psi_m }{2}sin\frac{\theta_m }{2}cos\frac{\phi_m }{2}}}},\\
&q_{2 m }={{\rm{cos\frac{\psi_m }{2}sin\frac{\theta_m }{2}cos\frac{\phi_m }{2} + sin\frac{\psi_m }{2}cos\frac{\theta_m }{2}sin\frac{\phi_m }{2}}}},\\
&q_{3 m} ={{\rm{sin\frac{\psi_m }{2}cos\frac{\theta_m }{2}cos\frac{\phi_m }{2}+ cos\frac{\psi_m}{2}sin\frac{\theta_m }{2}sin\frac{\phi_m }{2}}}}.
\end{align}
Note that in the above equations, we omit the time index $\tau$ for simplicity.

Therefore, the\emph{ measurement equation} can be expressed as
\begin{align}\label{equ:attitude177}
\mathbf z(\tau)=[q_{0 m}\ q_{1 m}\ q_{2 m}\ q_{3 m}]^T+\boldsymbol u,
\end{align}
where $\boldsymbol u$ is the measurement noise vector, and each element meets $\mathcal {CN}(0,\mathbf{Q}_{\boldsymbol \omega_k})$.

\begin{algorithm}[t]
\caption{: Attitude determination by Kalman filter}
\label{alg:algorithm1}
\begin{itemize}
\item \textbf{Step 1:} {Initialization: };
\item \textbf{Step 2:} {Prediction:} ${{ \mathbf q}}(\tau|\tau-1)=\boldsymbol \Gamma{{ \mathbf q}}(\tau-1)$;
\item \textbf{Step 3:} {Minimum prediction MSE:} $\boldsymbol\kappa(\tau|\tau-1)=\boldsymbol \Gamma(\tau-1)\boldsymbol\kappa(\tau-1)\boldsymbol \Gamma(\tau-1)^H+\mathbf{Q}_{\boldsymbol \chi}$;
\item \textbf{Step 4:} {Kalman gain matricx:}
$\mathbf\Upsilon(\tau)=\boldsymbol\kappa(\tau|\tau-1)[\boldsymbol\kappa(\tau|\tau-1)+\mathbf {Q}_{ \boldsymbol u}]^{-1}$;
\item \textbf{Step 5:} {Updation and normalization:} $\hat{{ \mathbf q}}(\tau)=\frac{{{ \mathbf q}}(\tau|\tau\!-\!1)\!+\!\mathbf\Upsilon(\tau)[\mathbf z(\tau)\!-\!{ \mathbf q}(\tau|\tau\!-\!1)]}{\left\{{{ \mathbf q}}(\tau|\tau\!-\!1)+\mathbf\Upsilon(\tau)[\mathbf z(\tau)\!-\!{ \mathbf q}(\tau|\tau\!-\!1)]\right\}^T\!\!\left\{{{ \mathbf q}}(\tau|\tau\!-\!1)\!+\!\mathbf\Upsilon(\tau)[\mathbf z(\tau)\!-\!{ \mathbf q}(\tau|\tau\!-\!1)]\right\}}$;
\item \textbf{Step 6:} {MMSE:} $\boldsymbol\kappa(\tau)=[\mathbf I-\mathbf \varpi_{k}(\tau)]\boldsymbol\kappa(\tau|\tau-1)$;
\item \textbf{Step 7:} Go to next time instant $\tau +1$.
\end{itemize}
\end{algorithm}
According to the system equation \eqref{equ:systemmodel} and the measurement equation \eqref{equ:attitude177}, the UAV attitude can be tracked by Kalman filter based data fusion \cite{EKF,UKF}, which integrates the outputs of the low-cost sensors to provide a precise attitude information. The detailed steps are shown in Algorithm~\ref{alg:algorithm1}.

For attitude adjustment, we need to further transform the quaternion to the attitude angle. Since $\mathbf{C}_b^n$ can be derived from the tracked quaternion as
\begin{align}\label{equ:attitude12}
{\bf{C}}_b^n({\hat{\bf{q}}}) &= {\boldsymbol\Xi ^{\rm{T}}}({\hat{\bf{q}}})\left[ {\begin{array}{*{20}{c}}
{{\hat{q}_4}{I_{3 \times 3}} - \left[ {\begin{array}{*{20}{c}}
0&{ - {\hat q_3}}&{{\hat q_2}}\\
{{\hat q_3}}&0&{ - {\hat q_1}}\\
{ - {\hat q_2}}&{{\hat q_1}}&0
\end{array}} \right]}\\
{ - {{\hat{\boldsymbol{\rho }}}^{\rm{T}}}}
\end{array}} \right]\notag\\&= \left[ {\begin{array}{*{20}{c}}
{{C_{11}}}&{{C_{12}}}&{{C_{13}}}\\
{{C_{21}}}&{{C_{22}}}&{{C_{23}}}\\
{{C_{31}}}&{{C_{32}}}&{{C_{33}}}
\end{array}} \right],
\end{align}
the attitude can be obtained from the definition of ${\bf{C}}_b^n({\hat{\bf{q}}})$ as
\begin{align}\label{equ:attitude16}
&\hat\psi(\tau)  = \arctan \left(\frac{{{C_{21}}}}{{{C_{11}}}}\right)=\psi(\tau)+\delta \psi(\tau),\\
&\hat\theta(\tau)  = \arcsin \left( - {C_{31}}\right)=\theta(\tau)+\delta \theta(\tau),\label{equ:attitude17}\\
&\hat\phi(\tau)  = \arctan \left(\frac{{{C_{32}}}}{{{C_{33}}}}\right)=\phi(\tau)+\delta \phi(\tau),\label{equ:attitude18}
\end{align}
where $\delta \psi(\tau),\ \delta \theta(\tau),\ \delta \phi(\tau)$ are the attitude errors due to the system error and measurement error, while $\psi(\tau),\ \theta(\tau),\ \phi(\tau)$ are the true values of the UAV attitudes.

\balance

\end{document}